\documentclass[pre,aps,twocolumn,showpacs,floatfix]{revtex4}
\usepackage{graphics}
\usepackage{graphicx}
\usepackage{epsf}
\begin{document}
\title{Non-equilibrium distributions at finite noise intensities}

\author{A.~Bandrivskyy, S.~Beri, D.~G.~Luchinsky$^\dag$}

\affiliation{Department of Physics, Lancaster University,
Lancaster LA1 4YB, UK}

\begin{abstract}
We analyse the non-equilibrium distribution in dissipative
dynamical systems at finite noise intensities. The effect of
finite noise is described in terms of topological changes in the
pattern of optimal paths. Theoretical predictions are in good
agreement with the results of numerical solution of the
Fokker-Planck equation and Monte Carlo simulations.
\end{abstract}

\pacs{05.40.+j, 02.50.-r, 05.20.-y}
\date{\today}
\maketitle 

\newcommand{\ud}{\mathrm{d}}
The analysis of the probability distribution in systems away from
thermal equilibrium is one of the main themes in statistical
physics~\cite{Onsager:53,Haken:75} which has recently attracted
renewed interest in the context of stochastic resonance
\cite{Nicolis:82} and stochastic ratchets \cite{Dykman:97a}.
Despite its importance, the theory of the probability
distributions in non-equilibrium systems is still in its infancy.

One of the keys to the solution of this problem is an
understanding of how a non-equilibrium system deviates far away
from a steady state~\cite{Onsager:53}. The probability of such
deviations is small and can be approximated by a Boltzmann-like
distribution of the form \cite{Ludwig:75a,Freidlin:84a}
\begin{equation}
\label{eq:WKB} \rho (x) = z(x) \exp\left({-S(x)/\epsilon}\right),
\quad \epsilon \to 0.
\end{equation}
Here the action function $S(x)$ plays the role of a
non-equilibrium potential~\cite{Graham:84a}, $z(x)$ is a prefactor
and $\epsilon$ is the noise intensity. The majority of earlier
studies have been restricted to the analysis of zero noise limit.
In this limit properties of $\rho(x)$ are determined by the
properties of optimal paths that are deterministic trajectories
along which the system moves in the course of large
deviations~\cite{Freidlin:84a,Ludwig:75a}. Accordingly, the
pattern of optimal paths plays a fundamental role in the
zero-noise-limit theory.

In ``real life'', however,  the noise intensity is always finite.
Calculations of $\rho(x)$ at finite noise intensity, that allow
for a direct comparison with an experiment, require analysis of
the all-important {\it prefactor}. Such analysis has recently been
initiated in a number of publications
\cite{Smelyanskiy:99a,Hanggi:00a,Maier:01a} dealing with the
closely related problem of noise-induced {\it escape} over an
unstable limit cycle.

The research has shown that one has ``to leave room for the
possibility that, even for small noise strengths $\epsilon$, more
than one'' optimal path contributes significantly to the
escape~\cite{Hanggi:00a,Maier:01a}. However, this plausible
assumption contradicts the fundamental properties of the pattern
of optimal paths including the existence of the singularities and
generic uniqueness of the most probable escape path (see
e.g.~\cite{Graham:84a,Jauslin:87a,Dykman:94a}). This inconsistency
does not allow the use of the methods of
\cite{Smelyanskiy:99a,Hanggi:00a,Maier:01a} to compute
noise-induced variations of the non-equilibrium distribution in
the {\it whole of phase space}. As mentioned in~\cite{Maier:01a}
the problem is ``intricate'' due to the subtle properties of the
pattern of optimal paths near the basin boundary.

In this Letter, the $\rho(x)$ is calculated over the whole of
phase space as a function of noise intensity. Specifically, we
address the intricate problem of the noise-induced changes in the
pattern of optimal paths near the singularities and show that
noise-induced variations of $\rho(x)$ are related to the shift of
the singularities and of the most probable escape paths. The
theoretical predictions are verified by Monte Carlo simulations
and the numerical integration of the Fokker-Planck equation (FPE).
In the boundary region we find the position of singularities
analytically, and provide a more accurate calculation of the
oscillating probability distribution discussed in the recent
publications \cite{Smelyanskiy:99a,Hanggi:00a,Maier:01a}, and
suggest an intuitively clear physical picture of these
oscillations.

To see why the computation of the prefactor calls for the analysis
of the noise-induced topological changes in the pattern of optimal
paths, let us consider a Langevin equation of the form
\begin{equation}
\dot{x}_i = K_i (x) + \sqrt{\epsilon}\sigma_{ij} \xi_j (t) \quad
i,j=1,2. \label{eq:langevin}
\end{equation}
Here $x_i$ are coordinates of a 2-dimensional dissipative system,
$\xi_j \left( t \right)$ are Gaussian zero-mean uncorrelated
sources of noise linearly mixed by a given noise matrix
$\sigma_{ij}$, and $\epsilon$ is a small parameter of the theory.

The evolution of the probability density $\rho (x,t)$ for the
system (\ref{eq:langevin}) is described by the  FPE
\begin{eqnarray}
\frac{\partial \rho}{\partial t}=\frac{\partial}{\partial x_i}
\left(-\rho K_{i}+ \frac{\epsilon}{2}\frac{\partial}{\partial
x_j} \left[ Q_{ij}\rho \right]\right), \label{eq:FPE}
\end{eqnarray}
where $Q = \sigma^T \sigma$ is a positively defined diffusion
matrix.

To the leading order of approximation in $\epsilon$, $S$
satisfies~\cite{Ludwig:75a,Freidlin:84a} a Hamilton-Jacobi
equation for a classical action in the form $ H (x,\nabla S)=0,$
where $H(x,p)$ is the Hamiltonian function equals
$\frac{1}{2}Q_{ij}p_i p_j + K_i(x) p_i$. The pattern of extreme
trajectories emanating from the stationary state of the system
(\ref{eq:langevin}) is found by integrating Hamiltonian equations
with appropriate initial conditions (see e.g. \cite{Ludwig:75a}),
and the action is found as an integral along these trajectories
\begin{eqnarray}
 \label{eq:HE}
 \dot x_i = K_i + Q_{ij}p_j, \quad
 \dot p_i = -\frac{\partial K_j}{\partial x_i}p_j,
 \quad \dot S = \frac{1}{2} Q_{ij} p_i p_j.
\end{eqnarray}
In general more than one trajectory can arrive at the same point
$x$~\cite{Dykman:94a}, but in the limit of $\epsilon \to 0$ the
only trajectories contributing to the $\rho(x)$ are those that
provide the global minimum of the action and form the pattern of
{\it optimal paths}.
\begin{figure}[!ht]
\center{\includegraphics[width=6cm,height=4cm]{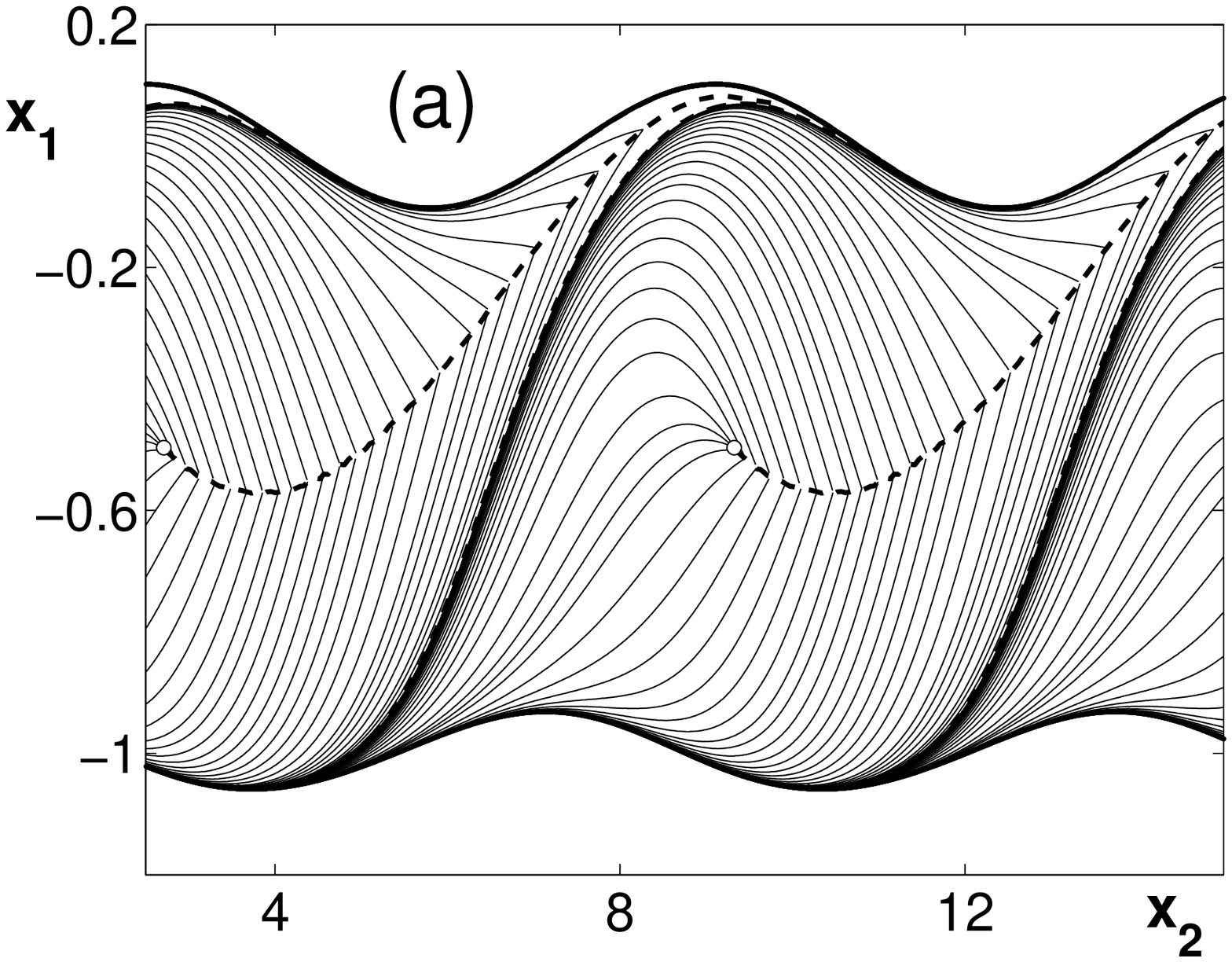}
\includegraphics[width=6cm,height=4cm]{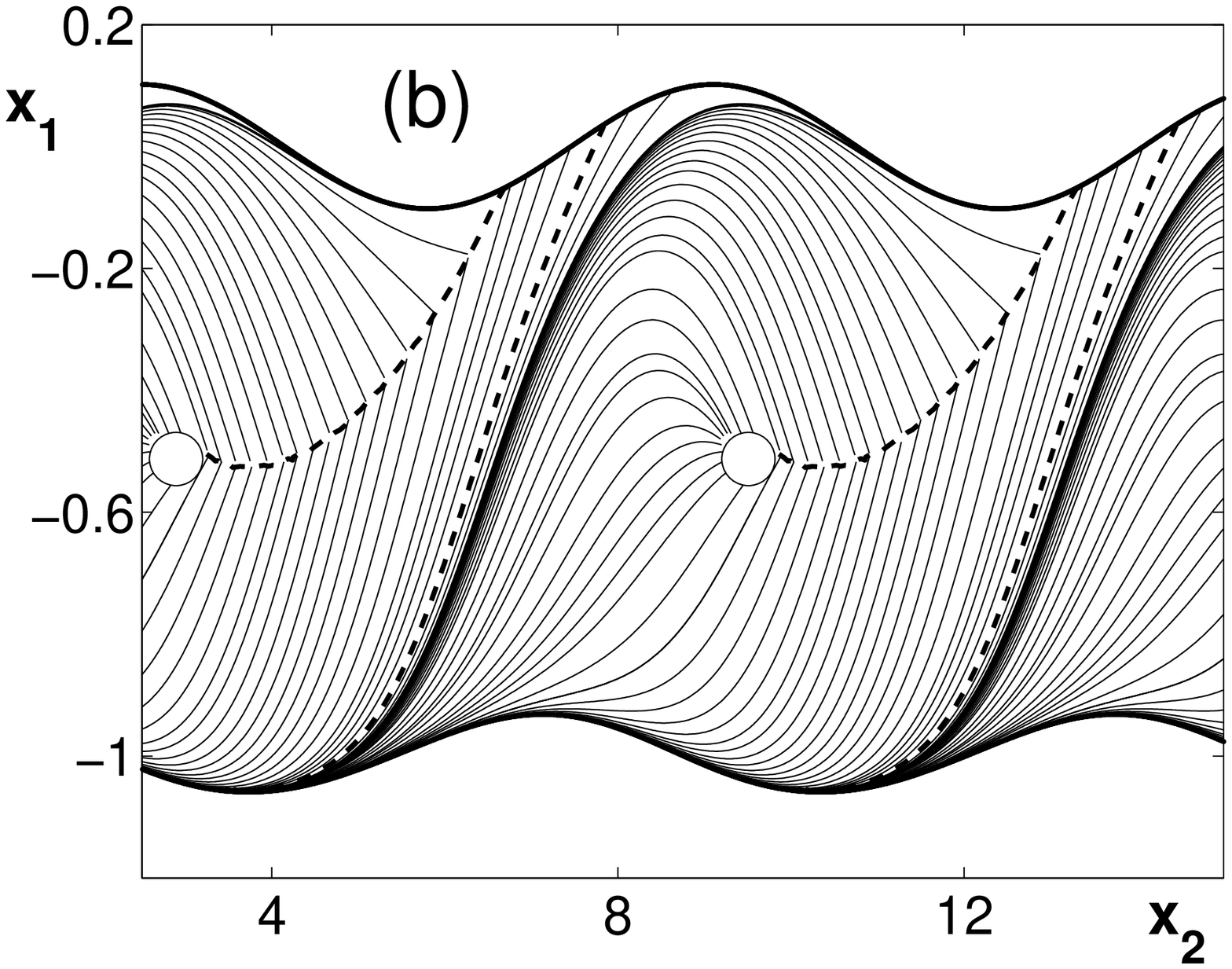}
} \caption{The part of the pattern of optimal paths between the
unstable and one of the stable limit cycles of a PDOD ($A=0.1,
\Omega=2\pi$) (a) for $\epsilon=0$ and (b) for $\epsilon = 0.015$.
The stable $x_s(t)$ and unstable $x_u(t)$ limit cycles are shown
by solid lines at the bottom and the top of the figure
correspondingly. The empty region inside the circle is the \lq\lq
breakdown region\rq\rq described in the text. The MPEP and the
switching line are indicated by dashed lines.} \label{figure1}
\end{figure}
As a result of global minimization the state space of
2-dimensional system (\ref{eq:langevin}) is separated by so-called
{\it switching lines} into areas to which the system arrives along
optimal paths of topologically different types~\cite{Dykman:94a}.
In particular, the escape from a metastable state of the system
(\ref{eq:langevin}) in the $\epsilon \to 0$ limit is governed by
the properties of the so-called {\it most probable escape path}
(MPEP) that emanates (at $t\to -\infty$) from a stationary state
of (\ref{eq:langevin}) and arrives (at $t\to +\infty$) at the
boundary of its basin of attraction. The pattern of optimal paths,
switching line, and the most probable escape paths are shown in
Fig.~\ref{figure1}(a) for a periodically driven overdamped motion
in a bistable Duffing (PDOD) system: $K_1 = x - x^3 + A \sin
\left( \Omega t \right)$, $K_2=\Omega=2\pi/T$,
$\sigma_{ij}=\sqrt{2}\delta_{i1}\delta_{j1}$ ($x_1=x$ and
$x_2=\Omega t$).

\begin{figure}[!ht]
\center {\includegraphics[width=6cm,height=3.6cm]{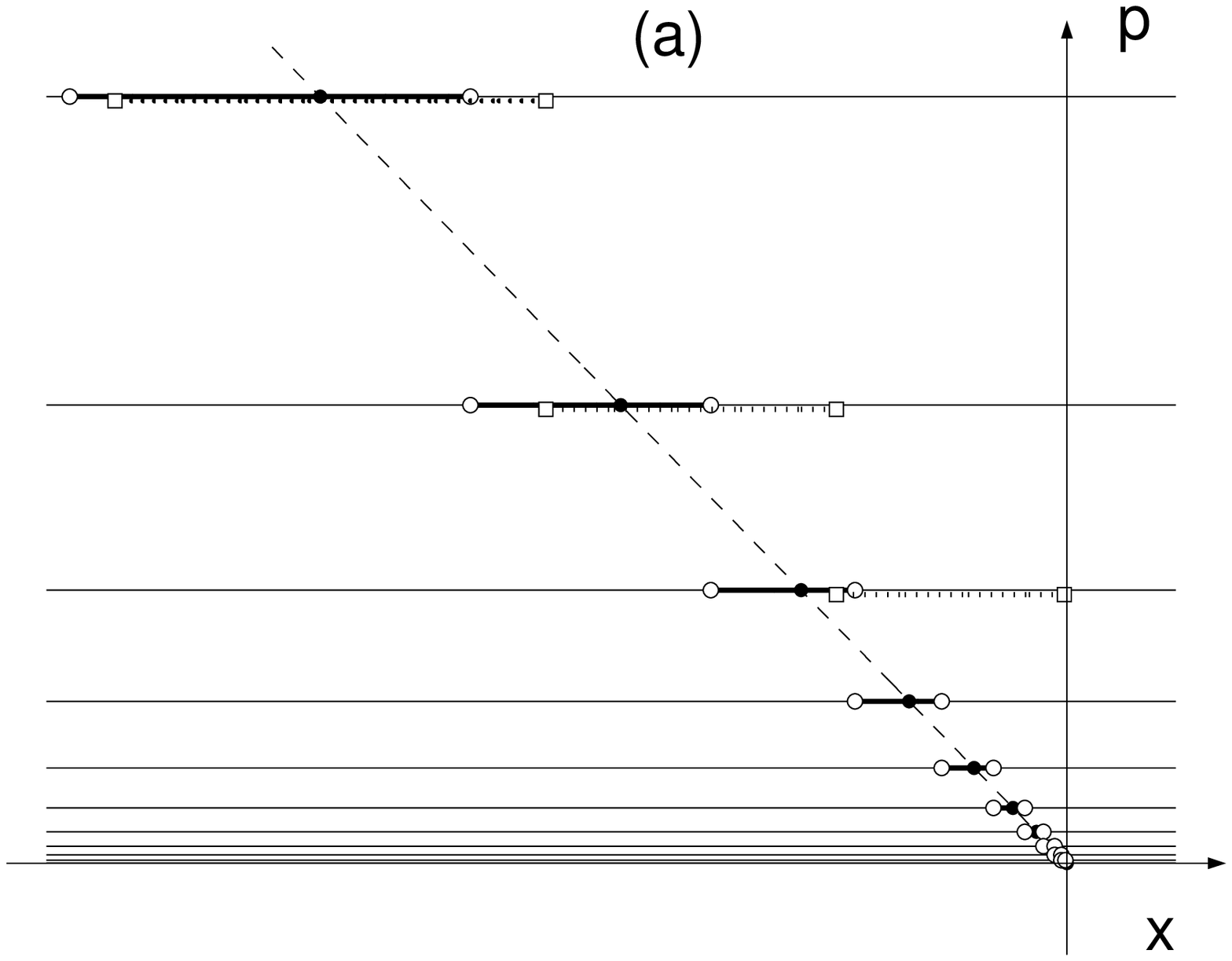}
\includegraphics[width=6cm,height=3.6cm]{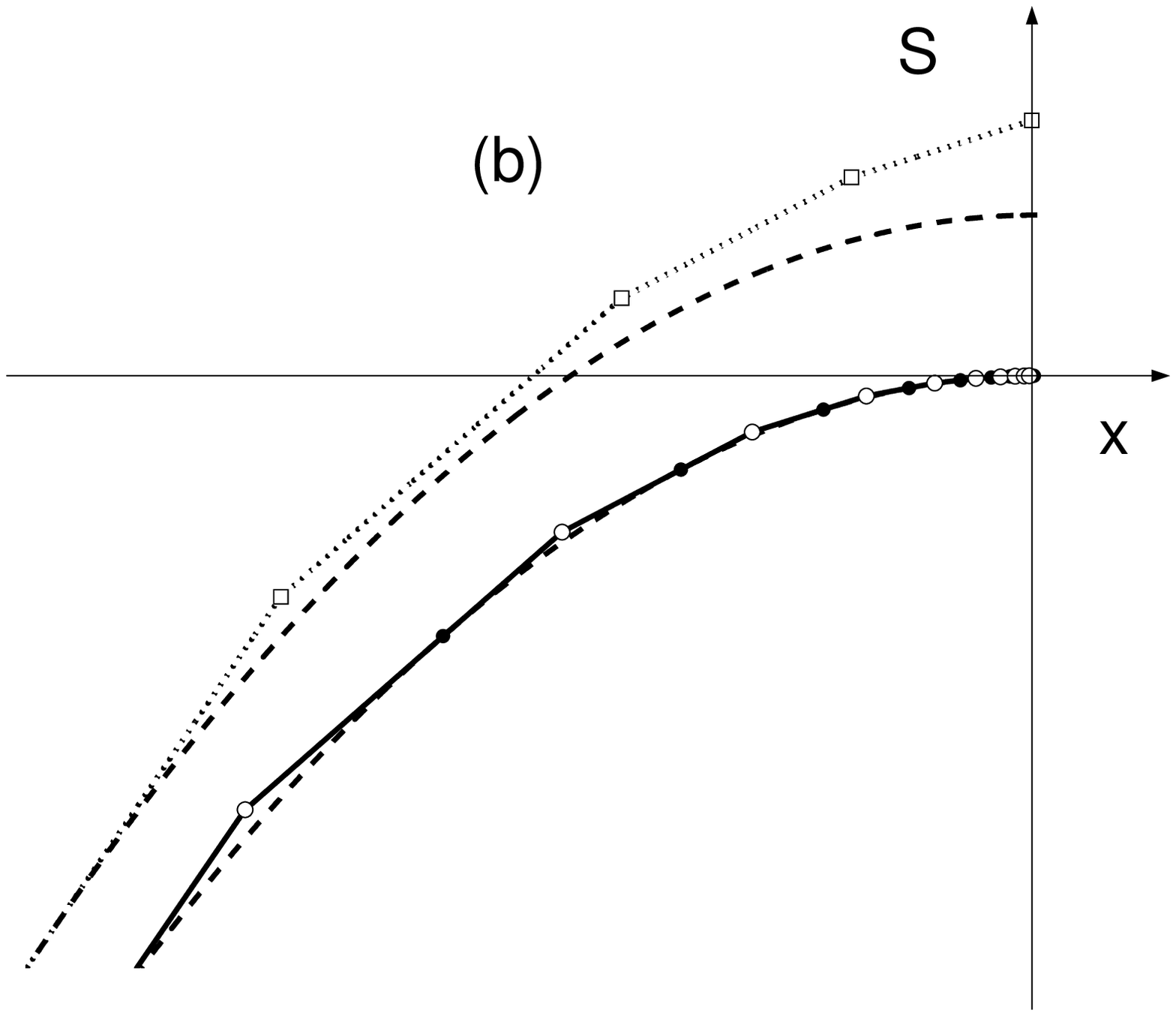}}
\caption{\label{figure2} (a) A section of the lagrangian
manifolds. ${\cal M}^{(s)}_u$ is shown by the dashed line. ${\cal
M}^{(u)}_s$ is a set of horizontal lines. Physically observable
part of the ${\cal M}^{(u)}_s$ formed by the optimal paths is
shown by the thick segments for $\epsilon=0$ and by dashed
segments for $\epsilon \neq 0$. The MPEP is shown by dots at the
intersection of the two manifolds. The switching line is shown by
the open circles for $\epsilon=0$ and by squares for $\epsilon
\neq 0$. (b) A section of the non-equilibrium potential (solid
line). The MPEP is shown by dots where $S(y)$ touches the parabola
(dashed line). The switching line corresponds to the points of
singularities of $S(y)$ (open circles). The section  of the
modified action surface $S-\epsilon \ln z$ is shown by dotted line
and shifted switching line (for $\epsilon \neq 0$) is shown by
squares.}
\end{figure}

The pattern is periodic in time and the behavior of the switching
line and of the MPEP near the basin boundary is described by the
scaling properties of the solution of the linearized Hamiltonian
equations
(c.f.~\cite{Smelyanskiy:99a,Hanggi:00a,Maier:01a,Smelyanskiy:97b})
in a Poincar\`{e} section (see Fig. \ref{figure2}) in the form
\begin{eqnarray}
 &&p_n = y_0^*\cdot W_0\cdot a^n, \label{eq:solution_1}\\
 &&y_n = y_0^*\cdot a^{-n}+p_n\cdot W_0, \label{eq:solution_2}
\end{eqnarray}
where $y_n = x(t) - x_u(t)$ is the distance to the limit cycle at
time $t=t_0+nT$, $a = e^{-\lambda T}$, $\lambda = 1/T
\int_{t}^{t+T}\partial_{x_1}K_1 dt'$, and $W_0 =
(1-a^2)\left[2\int_t^{t+T}a^2(t',t)dt'\right]^{-1}$ is the second
derivative of the action in the $x$ direction ($W_0 =
\partial_y\partial_y S$ see eq. (\ref{eq:prefactor_2}) below).

The first term in the RHS equation (\ref{eq:solution_2}) describes
the divergence of the zero-momentum trajectory from $x_u(t)$
corresponding to the motion on the unstable manifold of this cycle
${\cal M}^{(u)}_u$. The second term corresponds to the motion on
the stable manifold ${\cal M}^{(s)}_u$ of $x_u(t)$ with non-zero
momentum. We note that the initial position of the most probable
escape path $y_0^*=x(t_0) - x_u(t_0)$ is the only global parameter
of the linear problem. It defines the intersection of ${\cal
M}^{(s)}_u$ and the unstable manifold of the stable limit cycle
${\cal M}^{(u)}_s$ (see Fig. \ref{figure2}(a)). The latter is a
set of horizontal lines. The physically observable part of ${\cal
M}^{(u)}_s$ formed by the optimal paths is a set of segments (see
Fig. \ref{figure2}(a)) limited by the switching line.

The position of the switching line is found by noting that the
cross-section of the action $S(y)$, as a function of the distance
$y$ from the cycle, is a piecewise linear approximation of an
ideal parabola $S_{esc} -
\frac{1}{2}y^2W_0$~\cite{Graham:84a,Maier:01a,Smelyanskiy:97b},
where $S_{esc}$ is an escape action in zero-noise limit. The MPEP
corresponds to the points where $S(y)$ touches the parabola. The
switching line corresponds to the points of singularities of
$S(y)$.

It follows from this geometrical picture that at $\epsilon=0$: (i)
MPEP is the only optimal path that approaches the unstable limit
cycle at $t \to \infty$, (ii) the position of the switching line
scales as $y^{sl}_n=(a+1)y_0^*a^n/2$ and (iii) the switching line
also approaches the unstable limit cycle at $t \to \infty$ in
agreement with the qualitative picture shown in Fig.~1(a). A
similar picture can be obtained for the inverted van der Pol
oscillator (see~\cite{Maier:01a}(b)) and is generic for 2D systems
with unstable limit cycles.

In the presence of the finite noise intensity, this picture breaks
down. As we have mentioned above, one has to assume that, even for
small noise intensity, more than one trajectory contributes
significantly to the probability of crossing the
boundary~\cite{Hanggi:00a}. The choice of a {\it right} set of
trajectories that cross the boundary is a subtle
problem~\cite{Maier:01a} owing to the fact~\cite{Graham:84a} that
extreme trajectories intersect one another wildly near the
unstable limit cycle. For example, in~\cite{Smelyanskiy:99a} a
contribution to the escape from {\it all} the trajectories that
emanate from the steady state is estimated analytically.
In~\cite{Hanggi:00a,Maier:01a} it was suggested that the main
contribution to the escape comes from an {\it infinite discrete
set} of trajectories that are small perturbations of the MPEP,
which is found numerically. Neither of these choices is consistent
with the existence of the switching line, and the methods of
\cite{Smelyanskiy:99a,Hanggi:00a,Maier:01a} cannot be applied to
compute noise-induced variation of the $\rho(x)$ over the whole of
phase space (cf. \cite{Bandrivskyy:02}).

We therefore suggest that, to analyze the distribution in the
whole phase space, it is necessary to investigate the
noise-induced changes in the pattern of optimal paths. Such an
analysis can be conveniently performed in the next-to-leading
order of approximation in $\epsilon$ of the solution of the FPE,
yielding~\cite{Ludwig:75a} an equation for the prefactor in
(\ref{eq:WKB}) and for the Hessian matrix $S_{ij} \equiv
\partial_i\partial_j S$~\cite{Maier:97a}
\begin{eqnarray} &&\frac{dz}{dt} = - z \left(\partial_i
K_i + \frac{1}{2} Q_{ij} S_{ij}\right), \label{eq:prefactor_1} \\
&&\dot S_{ij}= - p_m\partial_i\partial_j K_m - S_{im}
\partial_j K_m \nonumber \\
&&  - S_{jm} \partial_i K_m -S_{jm}S_{ik}Q_{km}.
\label{eq:prefactor_2}
\end{eqnarray}

The second derivative of the action vanishes rapidly near $x_u(t)$
on ${\cal M}^{(u)}_s$, corresponding to the fact that ${\cal
M}^{(u)}_s$ is a set of horizontal lines near $x_u(t)$. Therefore
(for the PDOD) the noise induced correction to the action
$-\epsilon \ln(z(t))$ over one period can be estimated as
$\epsilon \lambda T$. As a result, a straightforward application
of the equations (\ref{eq:HE}), (\ref{eq:prefactor_1}),
(\ref{eq:prefactor_2}) leads to a discontinuity in the probability
distribution at the position of the switching line of  size
$\epsilon \lambda T$.

This discontinuity is unphysical. It can be removed if one defines
the position of a generalized switching line as a set of points:
$z_1\cdot\exp(-S_1/\epsilon)=z_2\cdot\exp(-S_2/\epsilon)$ for
which {\it probabilities} of arrival along topologically different
optimal paths are equal. We note that this is a natural
generalization of the original definition \cite{Dykman:94a} and
the latter is recovered when $\epsilon$ goes to zero. It has
physical significance of being the locus of experimentally
observable points where the curvature of the probability
distribution undergoes sharp qualitative changes (cf.
\cite{Luchinsky:97c}). The noise-induced changes in the physically
observable part of the ${\cal M}^{(u)}_s$ and in the $-\epsilon\ln
\rho(x)$ are shown in Fig. \ref{figure2}. Note the overall shift
of the $-\epsilon\ln \rho(x)$ due to the prefactor, and a stepwise
shifts of the linear segments by $\epsilon \lambda T$ without the
change of their slope. The geometrical analysis of these changes
gives the position of the modified switching line in the form
$y^{sl}_n = (a+1)y_n/2-\epsilon TW_0/((1-a)y_n)$.

The change in the position of the switching line has profound
consequences for the analysis of the non-equilibrium distribution
and the escape problem for the finite noise intensities. Firstly,
$\rho(x,t)$ calculated in the next to the leading order of WKB
approximation is now continuous everywhere in phase space.
Secondly, the modified switching line has to shift ``up'' (compare
Fig.~\ref{figure1} and Fig.~\ref{figure2}), and for the linear
regime it crosses the boundary at the point where the distance
from the original MPEP to the boundary is
$y^{(cr)}_n=\sqrt{2\epsilon TW_0/(1-a^2)}$. Since the new
switching line crosses the boundary, infinitely many optimal paths
can now cross the boundary. Moreover, the original MPEP ceases to
be the most probable escape path for a finite noise intensity,
since it is now ``cut off'' by the intersection with the new
switching line. Its role is taken by a different path that crosses
the boundary. Furthermore, the dominant contribution to the escape
from a metastable state is now governed by the continuous set of
trajectories which arrive to the boundary through a \lq\lq
corridor\rq\rq between the modified switching line and the
original MPEP. The modified pattern of optimal paths, including
the modified MPEP and the modified switching line, are shown in
 Fig.~\ref{figure1}(b). For small noise intensities, the set of
optimal trajectories that cross the boundary can be treated as
small perturbations of the original MPEP.

We note that the values of the prefactor $z$ and the Hessian
computed along the paths diverge as a trajectory approaches a
caustic. The approximation (\ref{eq:prefactor_1}),
(\ref{eq:prefactor_2}) breaks down here, and the method described
in this paper cannot be applied. The size of the breakdown region
scales as $O(\epsilon^{2/3})$ near caustic and as
$O(\epsilon^{3/4})$ near cusp (see e.g.
\cite{Schulman:81,Dykman:94a,Maier:97a}). Near the singularities
$\rho(x)$ can be calculated using the Maslov-WKB approximation
(see e.g. \cite{Dykman:94a,Maier:97a}). The results of these
calculations should merge in the \lq\lq far filed\rq\rq the
approximation found by the method described above. However, in
practice, when calculating $S-\epsilon\ln(z)$ using eqs.
(\ref{eq:HE}), (\ref{eq:prefactor_1}), (\ref{eq:prefactor_2}), the
problem of divergence can be avoided by partitioning the state
space into boxes with the size that scales as $\epsilon^{2/3}$ and
by assigning to each box the probability which is maximized over
the set of trajectories that pass through this box.
\begin{figure}[!ht]
\center{\includegraphics[width=6cm,height=3.7cm]{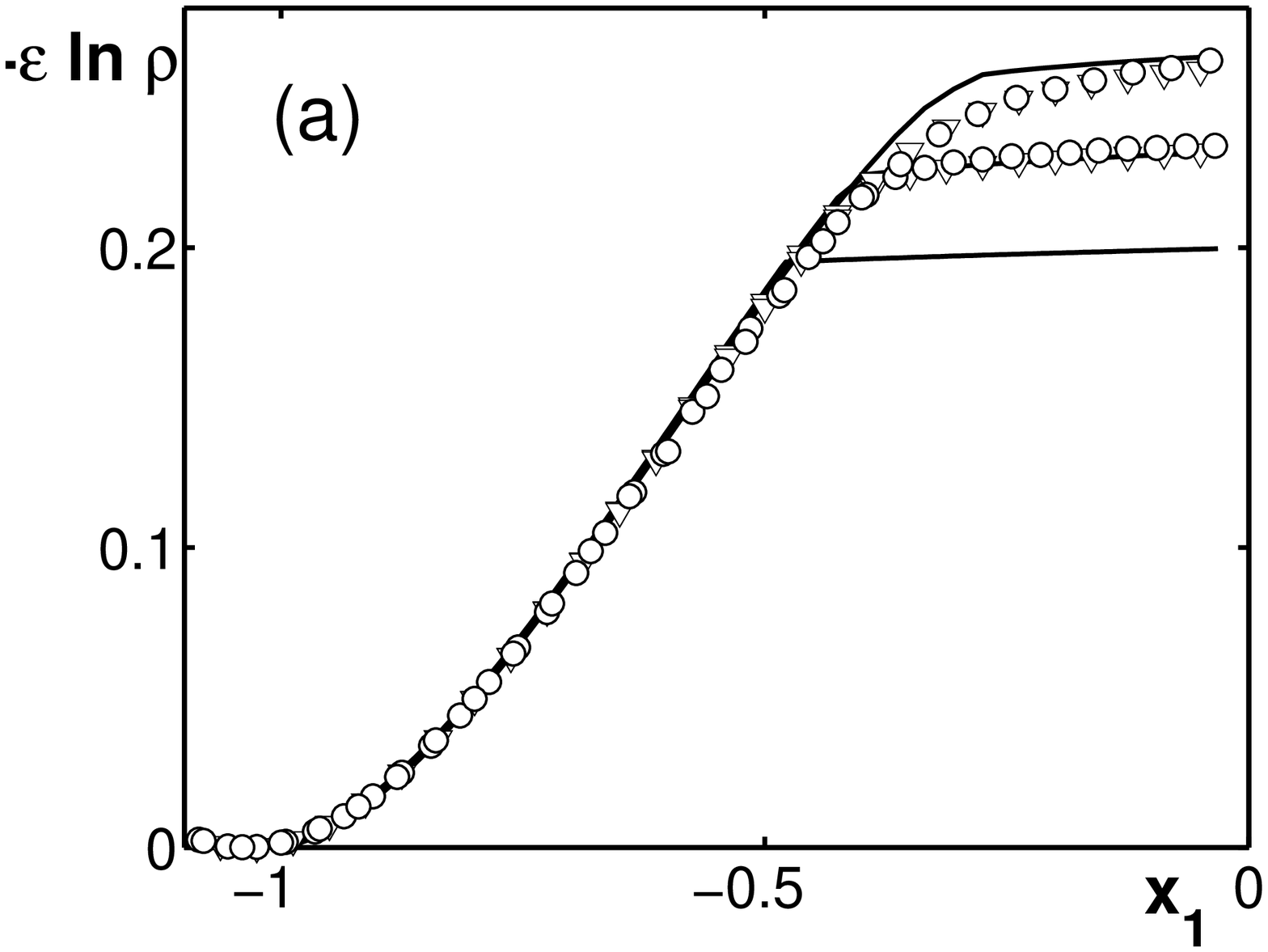}}
\center{\includegraphics[width=6cm,height=3.7cm]{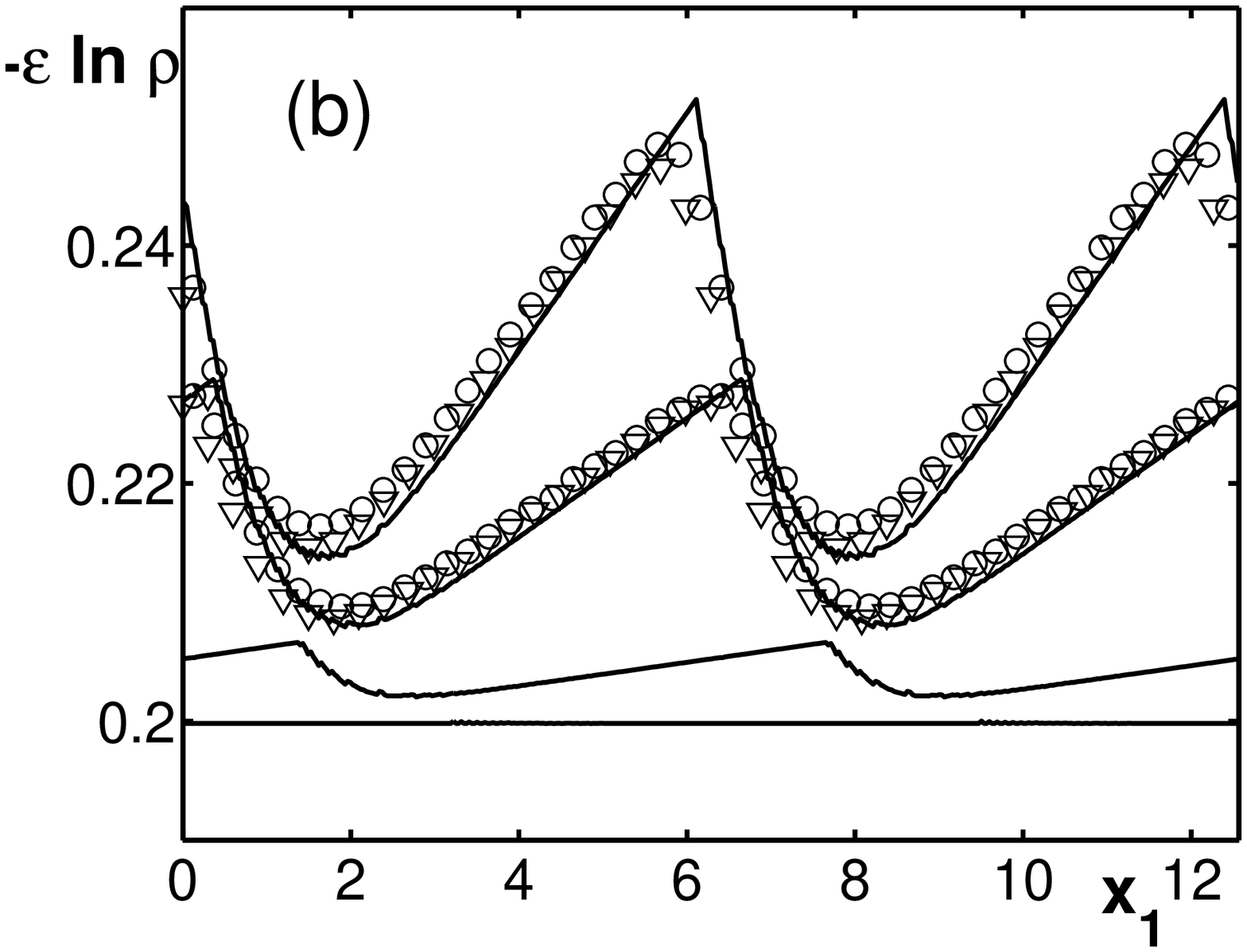}}
\caption{\label{figure3} (a) Sections at $x_1=4.1$ of the
$-\epsilon \ln \rho(x)$ calculated (solid lines) for different
values of $\epsilon$ (from the top to the bottom): 0.02, 0.01, 0.
The results of the Monte Carlo simulations (circles) and solution
of the FPE (triangles) for $\epsilon =$ 0.02 (top) and 0.01
(bottom) are shown for comparison. (b) The $-\epsilon \ln \rho(x)$
calculated (solid lines) along the unstable limit cycle $x^{(u)}$
for four values of $\epsilon$, from the top to the bottom, 0.01,
0.005, 0.001, and 0 are compared with Monte Carlo simulations
(circles) and numerical solution of the FPE (triangles) for
$\epsilon=$ 0.01 (top) and 0.005 (bottom) respectively.}
\end{figure}

To verify theoretical predictions we have performed Monte Carlo
simulations and numerical integration of the FPE for two model
systems: the PDOD and the inverted van der Pol oscillator
$K_1=x_2$, $K_2=-x_1- 2 \eta x_2 \left(1-x_1^2 \right)$,
$\sigma_{ij}=\sqrt{4 \eta}\delta_{i2}\delta_{j2}$. Qualitatively
similar behavior was obtained for both systems. In
Fig.~\ref{figure3}(a,b)  we compare theoretical predictions for
the function $S - \epsilon\ln(z)$ with the results of numerical
calculations for the PDOD in different cross-sections. Both
numerical techniques produce the same results and they are in good
quantitative agreement with the theory for a wide range of noise
intensity.

The behavior of $-\epsilon\ln (\rho(x))$ at the boundary on the
metastable state is shown in Fig.~\ref{figure3}(b) for various
values of the noise intensity. It can be seen that the amplitude
of the finite-noise induced oscillations increases with
$\epsilon$. The maxima of these oscillations correspond to the
intersection of the boundary with the modified switching line and
are the nondifferentiability points of the modified action surface
$S - \epsilon\ln(z)$. The minima of the distribution correspond to
the intersection of the boundary by the modified MPEP, and in the
linear regime are shifted with respect to the maxima by $\Delta t
\approx \ln[2\lambda T/(1-a^2_0)]/2\lambda$. For arbitrary values
of parameters the position of the modified switching line and MPEP
found in simulations coincide with those predicted theoretically.
As the noise intensity  $\epsilon$ decreases, the pattern of
optimal paths and the modified action approach their zero-noise
limits.

In conclusion, we have shown that the structure of singularities
of non-equilibrium potential in the zero-noise limit can be
modified in a self consistent way to include the effect of finite
noise intensity. As a result, a simple numerical scheme can be
used to calculate $\rho(x)$ over the whole of phase space in good
quantitative agreement with the results of simulations and
numerical integration of the FPE for two systems: inverted van der
Pol and periodically driven overdamped Duffing oscillators.

In the boundary layer we find analytically the position of the
modified switching line and modified MPEP, and give an intuitively
clear physical interpretation to the oscillations of the
probability at the boundary discussed in recent publications
\cite{Smelyanskiy:99a,Hanggi:00a,Maier:01a}. In particular, these
results  suggest that the oscillations of the escape rate as a
function of the noise intensity predicted in \cite{Maier:01a}
cannot be observed. The method suggested in this Letter can
readily be extended to higher dimensional systems.

The work was supported by the Engineering and Physical Sciences
Research Council (UK), the Joy Welch Trust (UK), the Russian
Foundation for Fundamental Science, and INTAS.

\end{document}